\pdfoutput=1
\documentclass[aps,prl,twocolumn,notitlepage,superscriptaddress,longbibliography,footinbib,preprintnumbers,floatfix]{revtex4-2}

\usepackage{times}
\usepackage{graphicx}
\usepackage{amsmath}
\usepackage{amssymb}
\usepackage{bm}
\usepackage{mathbbol}
\usepackage{xcolor}
\usepackage[colorlinks=true,citecolor=blue,linkcolor=blue,urlcolor=blue]{hyperref}
\usepackage{orcidlink}

\definecolor{runinred}{rgb}{0.75,0,0}
\newcommand{\runinsec}[1]{\textit{\textcolor{runinred}{#1.---}}\ }

\graphicspath{{figures/}}

\newcommand{\kk}{\bm{k}}
\newcommand{\ii}{\mathrm{i}}

\definecolor{ZZYcolor}{rgb}{0.1,0.5,0.4}

\begin{document}

\title{Layer-Asymmetry-Induced Topological Superconductivity in High-$T_c$ Bilayer Nickelates}

\author{Zhongyi Zhang\,\orcidlink{0000-0003-0887-9358}}
\email{zhongyizhang@ust.hk}
\affiliation{Department of Physics, Hong Kong University of Science and Technology, Clear Water Bay, Hong Kong, China}

\author{Congcong Le\,\orcidlink{0000-0003-2837-244X}}
\affiliation{RIKEN Interdisciplinary Theoretical and Mathematical Sciences (iTHEMS), Wako, Saitama 351-0198, Japan}

\author{Hoi Chun Po\,\orcidlink{0000-0002-5773-0712}}
\affiliation{Department of Physics, Hong Kong University of Science and Technology, Clear Water Bay, Hong Kong, China}

\author{Jiangping Hu\,\orcidlink{0000-0002-4837-7742}}\email{jphu@iphy.ac.cn}
\affiliation{New Cornerstone Science Laboratory,  Institute of Physics, Chinese Academy of Sciences, Beijing 100190, China}

\author{Xianxin Wu\,\orcidlink{0009-0004-3916-7120}}
\email{xxwu@itp.ac.cn}
\affiliation{CAS Key Laboratory of Theoretical Physics, Institute of Theoretical Physics, Chinese Academy of Sciences, Beijing 100190, China}

\author{Shengshan Qin\,\orcidlink{0009-0003-3484-8747}}
\email{qinshengshan@bit.edu.cn}
\affiliation{School of Physics, Beijing Institute of Technology, 100081 Beijing, China}

\begin{abstract}
The recent discovery of high-$T_{\rm c}$ superconductivity in bilayer nickelates has highlighted the crucial role of interlayer coupling and motivated proposals for dominant interlayer pairing. Whether such pairing can support topological superconductivity, however, remains unexplored. Here, we demonstrate that substrate-supported bilayer nickelate thin films can host an intrinsic time-reversal-invariant topological superconducting phase protected by diagonal mirror symmetry. Using a two-orbital model derived from first-principles calculations, we show that a modest layer-asymmetric potential and layer-staggered Rashba spin-orbit coupling drive the system into this topological phase over a broad parameter range with a dominant interlayer pairing.
The nontrivial topology arises from a reversal of the pairing sign on one of the spin-split bonding Fermi surfaces and is characterized by a nonzero mirror winding number along the diagonal direction, as well as a Majorana Kramers pair at the [11] edge.
The phase also features bulk gap nodes located slightly away from the diagonal directions, featuring chiral charge and zero-energy flat bands.
We discuss experimental routes for realizing and tuning this phase. Our results establish high-$T_{\rm c}$ bilayer nickelates as a promising platform for topological superconductivity, where tunable topological transitions and Majorana boundary modes can provide probes of unconventional pairing.
\end{abstract}

\maketitle

\runinsec{Introduction}The search for topological superconductivity is motivated by Majorana modes at boundaries or defects, whose existence is tied to the topology of the superconducting state~\cite{RevModPhys.83.1057,ChiuRMP2016,Alicea_2012,SatoAndo2017,10.1093/ptep/ptae065}.
Topological superconductivity can arise through two main routes, either from conventional pairing combined with suitable electronic states~\cite{FuKane2008,Sau2010,Zhang2024} or intrinsically from unconventional pairing~\cite{Kitaev2001,PhysRevB.61.10267}. The pairing structure is particularly important for time-reversal-invariant topological phases, because a noninteracting system in one or two dimensions cannot realize such a phase through proximity to a conventional $s$-wave superconductor~\cite{Haim2016NoGo}. Sign-changing pairing lifts this constraint and can give rise to such phases through its interplay with spin-orbit coupling (SOC) and multiband electronic structure~\cite{FuBerg2010,Deng2012TRISwave,ZhangKaneMele2013,Gaidamauskas2014}.

High-$T_{\rm c}$ superconductors are particularly appealing because their unconventional pairing and, in some materials, nontrivial band topology offer opportunities to realize topological superconductivity at elevated temperatures.
In iron-based superconductors, nontrivial band topology in proximity to bulk superconductivity produces Majorana zero modes trapped in vortices~\cite{hao2014topological,wu2016topological,wang2015topological,xu2016topological,QIN20191207}, which have been observed in experiments~\cite{ZhangSurface2018,WangMajorana2018,Kong2019, PhysRevX.8.041056,Machida2019,Li2022}.  Beyond this proximity-based mechanism, the sign-changing pairing of high-$T_{\rm c}$ superconductors can itself underpin intrinsic topological superconductivity. In cuprate superconductors, the nodal $d$-wave pairing can support spin-degenerate zero-energy flat bands on suitable edges~\cite{Hu1994,PotterLee2014}. In iron-based superconductors, $s_{\pm}$-wave pairing has been predicted to support time-reversal-invariant and higher-order topological phases with characteristic Majorana boundary modes~\cite{WuBoundary2020,Qin2022,ZhangDasSarma2021}.
Such intrinsic topological superconductivity driven by unconventional pairing has yet to be unambiguously established experimentally.
Its realization links Majorana physics directly to pairing symmetry and provides a phase-sensitive probe of the superconducting order~\cite{WuBoundary2020,Qin2022}, offering insight into the underlying pairing mechanism.

Bilayer nickelates provide a new setting in which to pursue this connection. Superconductivity has been discovered near $80$~K under pressure~\cite{Sun2023,ZhangZeroRes2024,WangBulk2024} and above $60$~K in strained films at ambient pressure~\cite{Ko2025,Zhou2025,Zhou2026}. The strong interlayer coupling brings both Ni $d_{x^2-y^2}$ and $d_{z^2}$ orbitals into the low-energy electronic structure~\cite{Luo2023,ZhangLin2023,Lechermann2023,Gu2025,Christiansson2023,LiuOptical2024,YangARPES2024,SunARPES2026}.
Distinct from cuprate and iron-based superconductors, an out-of-plane interlayer pairing is proposed as the dominant pairing component owing to strong antiferromagnetic interlayer coupling~\cite{YangSpm2023,Sakakibara2024,Gu2025,ShenDMRG2023,LiuApical2023,Lu2024,ZhangDagotto2024,OhZhang2023,Liao2023,Qu2024,YangZhangZhang2023,ZhangTrends2023,QinYang2023,Tian2024,LuSMG2026,Luo2024,Ryee2024,Xia2025,Zhan2025,Bejas2025,Le2025,Duan2025,Zhan2026}. This state generates sign-changing pairing on the bonding and antibonding Fermi surfaces in different mirror sectors, realizing a mirror-enforced $s_{\pm}$-wave pairing~\cite{ZhangQPI2026}.
However, this pairing belongs to the $A_{1g}$ representation, and in both pressurized bulk crystals, with space group $I4/mmm$, and films without substrate effects, with space group $P4/mmm$, symmetry indicators~\cite{Ono2020Indicators}, Fermi-surface formulas~\cite{ZhangFermi2026}, and even the possibly overcomplete classification based on the Atiyah--Hirzebruch spectral sequence~\cite{Ono2024Classification} find no nontrivial phase.
Topologically nontrivial superconductivity in bilayer nickelates must therefore arise from other effects.
Thin films offer additional tunability through substrate-induced strain and charge transfer, which modify the crystal and electronic structures~\cite{Ko2025,Zhou2025,Le2025}, as well as through broken layer symmetry~\cite{Geisler2025,Zhao2025,bhatt2026structural,Le2025,FanVishwanath2025,OhZhangPDW2025} and spin-orbit splittings. Together with the sign-changing interlayer pairing, these features motivate us to explore intrinsic topological superconductivity in bilayer nickelates.

\begin{figure}[!tp]
 \centering
 \includegraphics[width=0.98\columnwidth]{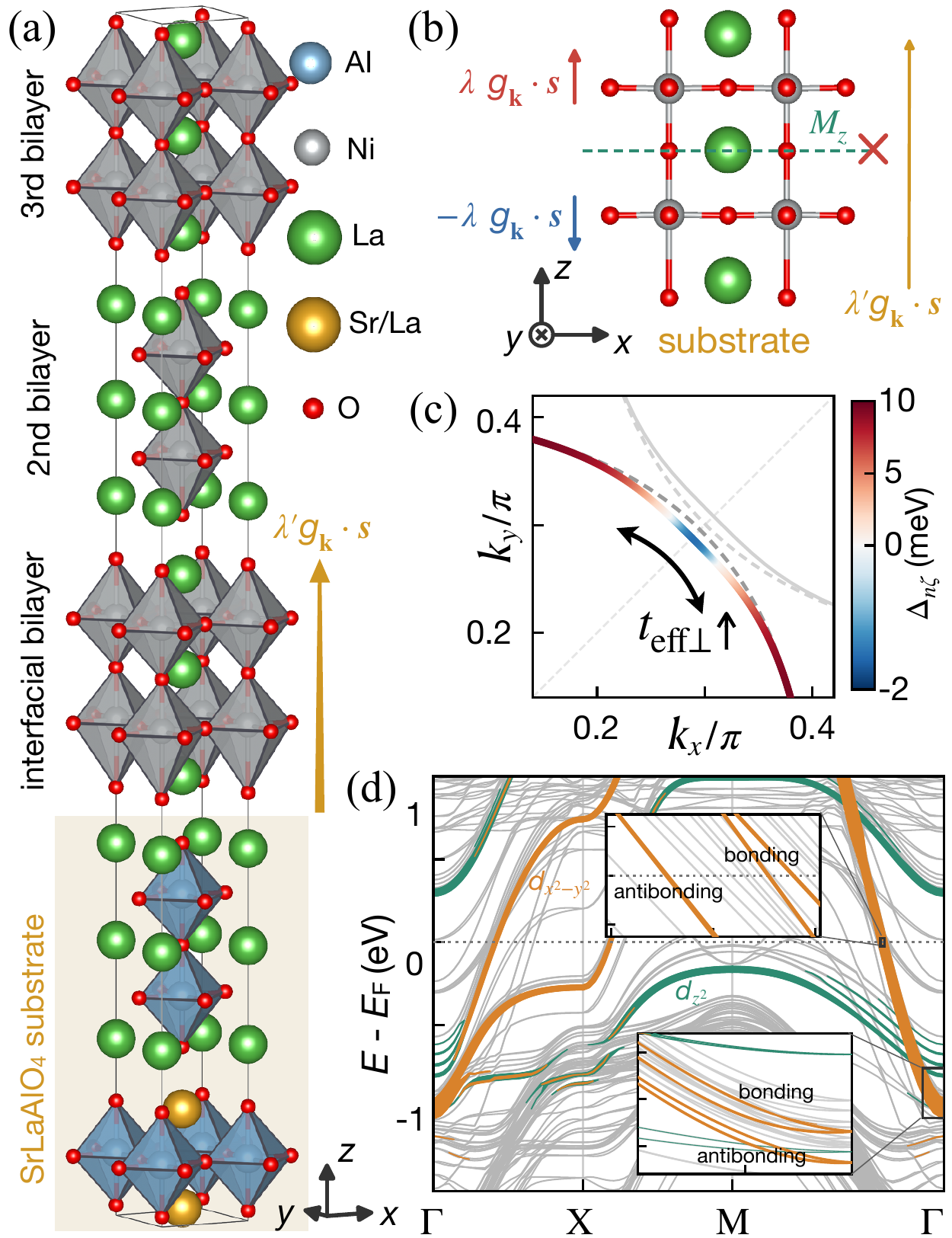}
 \caption{\label{figDFT}(a) Schematic of a La$_3$Ni$_2$O$_7$ film grown on SrLaAlO$_4$ with a LaAlO$_3$ layer (shaded), showing the interfacial bilayer and the two bilayers above it. The gold arrow marks the interface electric field, which sets the layer-uniform Rashba coupling and weakens away from the interface.
(b) Side view of the interfacial bilayer.
The red and blue arrows mark the opposite polarities of the layer-staggered Rashba coupling ($\pm\lambda$) on the top and bottom layers, and the gold arrow marks the layer-uniform Rashba coupling ($\lambda'$) shared by both layers.
The dashed line is the mirror plane $M_z$, broken by the substrate.
(c) Bonding $\chi=+$ Fermi surface of the $\alpha$ pocket colored by the intraband pairing.  On the diagonal (dashed), the pairing equals $\Delta_{n\zeta}$ in Eq.~\eqref{eq:projected} with $n=\zeta=+$. The parameters are chosen as $V=27$~meV, $\lambda=21.6$~meV, $\Delta_\parallel=-5$~meV, and $\Delta_\perp=15$~meV.
The double arrow marks the increase of the effective interlayer hopping $t_{{\rm eff}\perp}$ away from the diagonal direction.
(d) Bands with SOC of the substrated film, with the second bilayer $d_{x^2-y^2}$ (orange) and $d_{z^2}$ (green) weights shown by the line width. The upper inset enlarges the bonding and antibonding bands near the Fermi level along $\Gamma$-$\mathrm{M}$. The lower inset enlarges the bands near $\Gamma$ along $\Gamma$-$\mathrm{M}$, where SOC splits both the bonding and antibonding $d_{x^2-y^2}$ bands.}
\end{figure}

In this work, we demonstrate that high-$T_{\rm c}$ bilayer nickelate films can host an intrinsic time-reversal-invariant nodal topological superconductivity with an odd mirror winding number, by combining first-principles calculations and effective-model analysis. We find that the substrate naturally provides the effects needed for this phase. Based on the band structure of La$_3$Ni$_2$O$_7$ films on substrate, we construct an effective two-orbital model incorporating a substrate-induced layer-asymmetric potential, layer-staggered and uniform Rashba SOC terms.
With a dominant interlayer pairing, the naturally arising, modest layer potential and layer-staggered SOC drive the system into a topological superconducting phase in class DIII over a large parameter window.
The nontrivial topology originates from a pairing-sign reversal on one of the spin-split bonding Fermi surfaces and gives rise to a Majorana Kramers pair on the [11] edge. Away from the diagonals, orbital hybridization enhances the effective interlayer hopping and restores the pairing sign in the presence of layer asymmetry, producing bulk nodes slightly deviating from the diagonal directions. These nodes carry chiral charges and give rise to zero-energy flat bands on the [21] edge. Our results establish high-$T_{\rm c}$ bilayer nickelates as a promising platform for topological superconductivity, where tunable topological transitions and Majorana boundary modes offer novel probes of the pairing symmetry.

\runinsec{Electronic structure and effective model}
We begin by analyzing the electronic structure of La$_3$Ni$_2$O$_7$ thin films grown on a SrLaAlO$_4$ substrate, as sketched in Fig.~\ref{figDFT}(a). For NiO$_2$ bilayers near the interface, the low-energy states are dominated by the $d_{z^2}$ and $d_{x^2-y^2}$ orbitals, as shown in Fig.~\ref{figDFT}(d). The interface dipole breaks the mirror symmetry and modifies the carrier density~\cite{Nakagawa2006,Le2025}. It also introduces a layer-asymmetric potential and an additional Rashba SOC. The inset of Fig.~\ref{figDFT}(d) shows an enlarged view of the bands along $\Gamma$-$\mathrm{M}$, revealing spin splitting of both the bonding and antibonding $d_{x^2-y^2}$ bands in the bilayer near the interface.

We construct an effective model to capture the band structure from first-principles calculations. In the absence of SOC and substrate effects, the low-energy electronic structure is described by a two-orbital tight-binding Hamiltonian,
\begin{equation}
 \begin{aligned}
 h_{0}(\kk)={}&(\varepsilon_+\ell_0+t_+\ell_x)\sigma_0
             +(\varepsilon_-\ell_0+t_-\ell_x)\sigma_z\\
 &+(\varepsilon_{xz}\ell_0+t_{\perp xz}\ell_x)\sigma_x.
 \end{aligned}
 \label{eq:model1}
\end{equation}
Here, $\ell_i$ and $\sigma_i$ are Pauli matrices in the layer and orbital space respectively. In Eq.~\eqref{eq:model1}, the terms $\varepsilon_\pm=(\varepsilon_x\pm\varepsilon_z)/2$ depict the intraorbital hoppings between different neighbour sites in the same layer, $t_\pm=(t_{\perp x}\pm t_{\perp z})/2$ describe intraorbital hoppings between different layers, and $\varepsilon_{xz}$ and $t_{\perp xz}$ correspond to the intra- and interlayer hybridizations between the two orbitals respectively. The explicit expressions and tight-binding parameters are provided in the Supplemental Material (SM).

We next incorporate SOC and substrate-induced mirror-symmetry-breaking effect. Although an isolated bilayer is globally inversion-symmetric, each NiO$_2$ layer lacks local inversion symmetry, allowing a layer-staggered Rashba SOC~\cite{ZhangZunger2014,YuanZunger2019,Qin2022,WuBiS2_2017,AgterbergFeSe2017,FischerReview2023,ZhangLiuHSP2020}. As illustrated in Fig.~\ref{figDFT}(b), the inversion center and the mirror plane $M_z$ lie midway between the two NiO$_2$ layers. Consequently, the Rashba SOC has equal strength and opposite sign in the two layers. Symmetry therefore permits the term
$h_{\rm SOC}(\kk)=\lambda\bm g_{\kk}\cdot\bm s\,\ell_z$,
where $\bm g_{\kk}=(-\sin k_y,\sin k_x,0)$ and $\bm s$ denotes the Pauli matrices in spin space. This term preserves the twofold degeneracy at each $\kk$ enforced by combined inversion and time-reversal symmetry, while producing opposite spin textures in the two layers. Such layer-resolved spin textures can be probed by spin-resolved angle-resolved photoemission spectroscopy (ARPES)~\cite{WuBiS2_2017}.
The substrate breaks $M_z$ through an interface electric dipole, generating a layer-asymmetric potential that we model as $h_{\rm sub}=V\ell_z$. The dipole also permits a layer-uniform Rashba term,
$h'_{\rm SOC}(\kk)=\lambda'\bm g_{\kk}\cdot\bm s\,\ell_0$,
whose strength is set by the interface electric field. For simplicity, we neglect the orbital dependence of these terms.

The substrate-induced parameters $V$ and $\lambda'$ depend on the distance from the interface and are expected to decrease for NiO$_2$ bilayers farther from it. For the bilayer near the interface, first-principles calculations reveal different spin splittings of the bonding and antibonding bands along $\Gamma$-$\mathrm{M}$, consistent with finite $V$, $\lambda$, and $\lambda'$. Fitting the effective model to these bands yields $V=27$~meV, $\lambda=21.6$~meV, and $\lambda'=10$~meV. The intrinsic SOC strength $\lambda$ is close to its bulk value and fitting details are provided in the SM.

We now add superconducting pairing to the normal-state Hamiltonian $h_{\rm N}=h_0+h_{\rm SOC}+h_{\rm sub}$, where we neglect $h'_{\rm SOC}$ because this layer-uniform term only slightly shifts the Fermi surfaces and does not change the topology (see Discussion). Theoretical studies suggest that strong interlayer coupling favors dominant interlayer pairing, accompanied by a smaller intralayer component~\cite{Gu2025,Lu2024,Luo2024}. In the Nambu basis
$\psi^\dagger(\kk)=(c^\dagger(\kk),\,\ii s_yc(-\kk))$,
the Bogoliubov--de Gennes (BdG) Hamiltonian takes the form
\begin{equation}
 H_{\mathrm{BdG}}(\kk)=h_{\rm N}(\kk)\tau_z+\Delta\tau_x,
 \label{eq:bdg}
\end{equation}
where $\tau_i$ are Pauli matrices in Nambu space and
$\Delta=\Delta_\parallel\ell_0+\Delta_\perp\ell_x$.
The parameters $\Delta_\parallel$ and $\Delta_\perp$ denote the on-site intralayer and interlayer spin-singlet pairing amplitudes, respectively. In the bonding/antibonding basis of the mirror-symmetric bilayer, the corresponding pairing amplitudes are $\Delta_\parallel+\Delta_\perp$ and $\Delta_\parallel-\Delta_\perp$.
We choose $\Delta_\parallel=-5$~meV and $\Delta_\perp=15$~meV, giving gap magnitudes of $10$~meV and $20$~meV on the $\alpha$ and $\beta$ Fermi pockets, respectively, consistent with recent experimental observations~\cite{WangSTM2026}.
Finally, substrate-induced mirror-symmetry breaking also allows different intralayer pairing amplitudes in the two layers. However, this asymmetry enters as a higher-order perturbative correction associated with $h_{\rm sub}$ and is therefore neglected in the following analysis.

\runinsec{Topological mirror superconductivity on $\Gamma$-$\mathrm{M}$}
We turn to the topological property of the superconductivity in substrated La$_3$Ni$_2$O$_7$ described by Eq.~\eqref{eq:bdg}. The system respects both the time-reversal symmetry $\mathcal T=\ii s_yK$ and particle-hole symmetry $\mathcal C=s_y\tau_yK$ where $K$ is the complex conjugation operation, placing it in class DIII according to the Altland--Zirnbauer classification~\cite{AltlandZirnbauer1997,ChiuRMP2016}. In the absence of the substrate, La$_3$Ni$_2$O$_7$ thin film follows the space group $P4/mmm$, and the pairing state in Eq.~\eqref{eq:bdg} belongs to the $A_{1g}$ irreducible representation. And by considering the symmetry constraints, it can be verified the pairing state is thoroughly topologically trivial~\cite{Ono2020Indicators,ZhangFermi2026,Ono2024Classification}. When the substrate is taken into account, encoded with the layer-staggered Rashba SOC, as we shall show the La$_3$Ni$_2$O$_7$ thin film can be driven into the topological superconducting state.

We first focus on the line $k_x=k_y$, i.e., the $\Gamma$-$\mathrm{M}$ line, where the mirror symmetry exchanging $k_x$ and $k_y$ is preserved even in the presence of the substrate. Under the basis for Eq.~\eqref{eq:bdg}, the above mirror symmetry takes the matrix form $M_{x\bar{y}}=-\sigma_z\exp(-\ii\pi\,{\hat{\bm n}\cdot\bm s}/2)$, with $\hat{\bm n}=(-1,1,0)/\sqrt2$ the normal of the mirror plane. The whole system can be decoupled according to $M_{x\bar{y}}$ along the line $k_x=k_y$. Moreover, the hybridization between the $d_{x^2-y^2}$ and $d_{z^2}$ orbitals  vanishes, i.e., $\varepsilon_{xz} = t_{\perp xz} = 0$ in Eq.~\eqref{eq:model1}, due to their different symmetries with only the $d_{x^2-y^2}$ orbital left on the Fermi level on $k_x=k_y$ as shown in Fig.~\ref{figDFT}(d). The decoupled subspaces can be labeled by the mirror eigenvalues of the $d_{x^2-y^2}$ orbital $\ii \zeta$ with $\zeta = \pm$. By setting $k_x = k_y = k$, we can obtain the normal-state Hamiltonian in each $M_{x\bar{y}}$-invariant sector
\begin{equation}
 h_\zeta(k)=\xi(k)\ell_0+t_\perp(k)\ell_x+[V+\zeta\,r(k)]\ell_z ,
 \label{eq:mirror-bdg}
\end{equation}
where $r(k)=\sqrt2\lambda\sin k$, $\xi(k)=\varepsilon_x(k,k)$ and $t_\perp(k)=t_{\perp x}(k,k)$.  The BdG Hamiltonian in the corresponding mirror sector reads $H_{\mathrm{BdG}, \zeta} = h_\zeta\tau_z + \Delta\tau_x$.

\begin{figure}[bp]
 \centering
 \includegraphics[width=0.98\columnwidth]{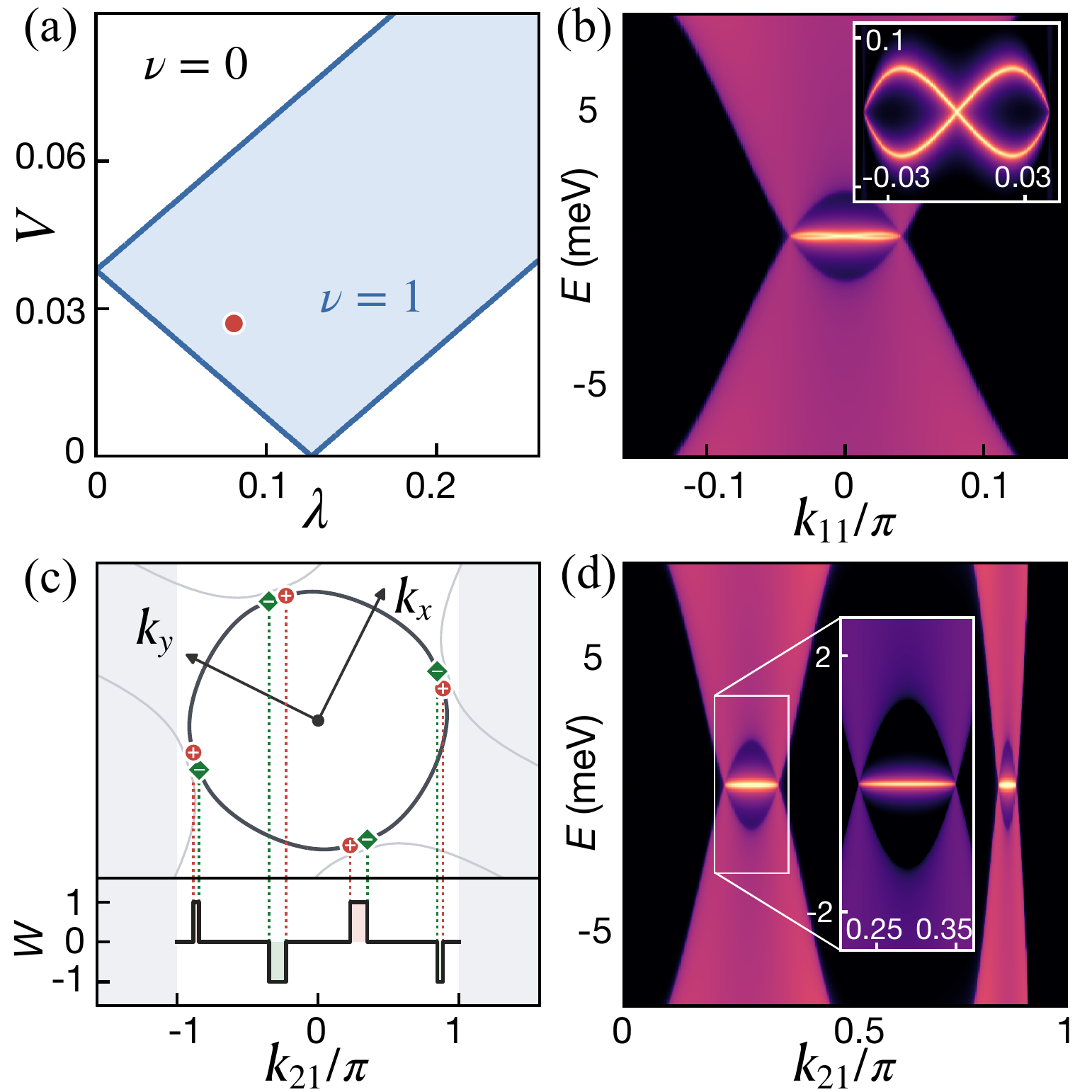}
 \caption{\label{figwindow}(a) Topologically nontrivial region of Eq.~\eqref{eq:straddle}.
 The horizontal scale is $a\lambda$ in eV\,\AA\ and the vertical scale is $V$ in eV. The red dot marks the representative parameters used in this work, $V=27$~meV and $\lambda=21.6$~meV ($a\lambda=81$~meV\,\AA).
(b) Superconducting edge modes on the [11] edge, with the conserved momentum $k_{11}=k_x-k_y$.
(c) Brillouin zone in the frame of a [21] edge (top) and winding number $W$ at fixed $k_{21}=k_x-2k_y$ (bottom).  Circles and diamonds are the nodes with chiral charges $Q=+1$ and $-1$ on the bonding $\chi=+$ Fermi surface (dark curve).
(d) Same as (b) for [21] edge, where flat band exists at $k_{21}$ with nonzero winding number in (c).}
\end{figure}

As mentioned, the system respects the time-reversal symmetry and particle-hole symmetry, the product of which gives the chiral symmetry $\mathcal S=-\ii\mathcal T\mathcal C=\tau_y$ anticommuting with $H_{\rm BdG}$. Moreover,  along $\Gamma$-$\mathrm{M}$, it satisfies $[\mathcal S, M_{x\bar{y}}] = 0$ for the pairing in Eq.~\eqref{eq:bdg}. Thus, in each $M_{x\bar{y}}$ subspace the chiral symmetry maintains and we can consider its winding number. Utilizing the eigenstates of the chiral symmetry operator, one can transform $H_{\mathrm{BdG}, \zeta}$ into the off-diagonal form, with the off-diagonal block being $q_\zeta=h_\zeta-\ii \Delta$. The winding number in the mirror sector $\ii \zeta$ on $\Gamma$-$\mathrm{M}$ can be calculated as
\begin{equation}
    w_\zeta = \int_{-\pi}^{\pi} \frac{\mathrm{d}k}{2\pi\ii} \,\partial_k\log z_\zeta
	= \int_{-\pi}^{\pi} \frac{\mathrm{d}k}{2\pi\ii} \,\partial_k\log \det q_\zeta,
	\label{eq:winding}
\end{equation}
with $\det q_\zeta = (\xi-\ii \Delta_\parallel)^2-(t_\perp-\ii \Delta_\perp)^2-(V+\zeta r)^2$. The time-reversal symmetry exchanges the two mirror subspaces demanding $w_+=-w_-$, and the whole system on the $k_x = k_y$ line can be characterized by the mirror winding number $w = ( w_+ - w_- )/2 = w_+$. It is worth mentioning that in general along the $k_x = k_y$ line the system can be viewed as a 1D system belonging to class DIII which is featured with a $\mathbb Z_2$ topological index according to the Altland--Zirnbauer classification~\cite{AltlandZirnbauer1997,ChiuRMP2016}. The $\mathbb Z_2$  index is actually equal to the parity of the mirror winding number, i.e., $(-1)^\nu=(-1)^{w}$~\cite{QiHughesRaghuZhang2009}.

Based on the above formula, the topological superconducting phase diagram with respect to $V$ and $\lambda$ can be straightforwardly achieved by solving $z_\zeta(k)=0$ in Eq.~\eqref{eq:winding}. To be specific, the imaginary part of the above equation is $\xi(k)\Delta_\parallel=t_\perp(k)\Delta_\perp$. It determines the momenta where the bulk energy gap closes.  For the band structure and pairing orders in our consideration, we get one solution pair $\pm k_0$.  The real part of $z_\zeta(k)=0$ turns out to be $\xi^2 - t_\perp^2 + \Delta_\perp^2 - \Delta_\parallel^2 = (V+\zeta r)^2$. Substituting $\pm k_0$ into the equation, one can get the topological superconducting state when it satisfies
\begin{equation}
	\begin{gathered}
		|V-r_0|<A_{\rm c}<V+r_0,
	\end{gathered}
	\label{eq:straddle}
\end{equation}
where
\begin{equation}
	\begin{gathered}
		A_{\rm c}^2 = t_\perp^2(k_0)\left( \frac{\Delta_\perp^2}{\Delta_\parallel^2} - 1 \right)
		+ \left( \Delta_\perp^2-\Delta_\parallel^2 \right),
	\end{gathered}
	\label{eq:straddle1}
\end{equation}
with $t_\perp(k_0)=t_{\perp x}(k_0,k_0)$ and $r_0=r(k_0)$. Here, we assume positive $V$ and $\lambda$, and take use of the fact $r(-k) = -r(k)$ and $t_\perp(-k) = t_\perp(k)$. Using the above parameters for La$_3$Ni$_2$O$_7$, we plot the topological superconducting phase diagram in Fig.~\ref{figwindow}(a). For the substrate effect and the staggered Rashba SOC fit from first-principles calculations in the above, i.e. $V=27$~meV and $\lambda=21.6$~meV, La$_3$Ni$_2$O$_7$ lies in the topological superconducting phase with a gap minimum about $1.3$~meV along $\Gamma$-$\mathrm{M}$. And it has a Majorana Kramers pair at the projecting point of  $\Gamma$-$\mathrm{M}$ on [11] edge as shown in Fig.~\ref{figwindow}(b). We point out that the Majorana Kramers pair could survive even when the mirror symmetry $M_{x\bar{y}}$ is broken, provided that the $\mathbb Z_2$ index protected by the time-reversal symmetry stays unchanged.

\begin{figure}[bp]
 \centering
 \includegraphics[width=0.95\columnwidth]{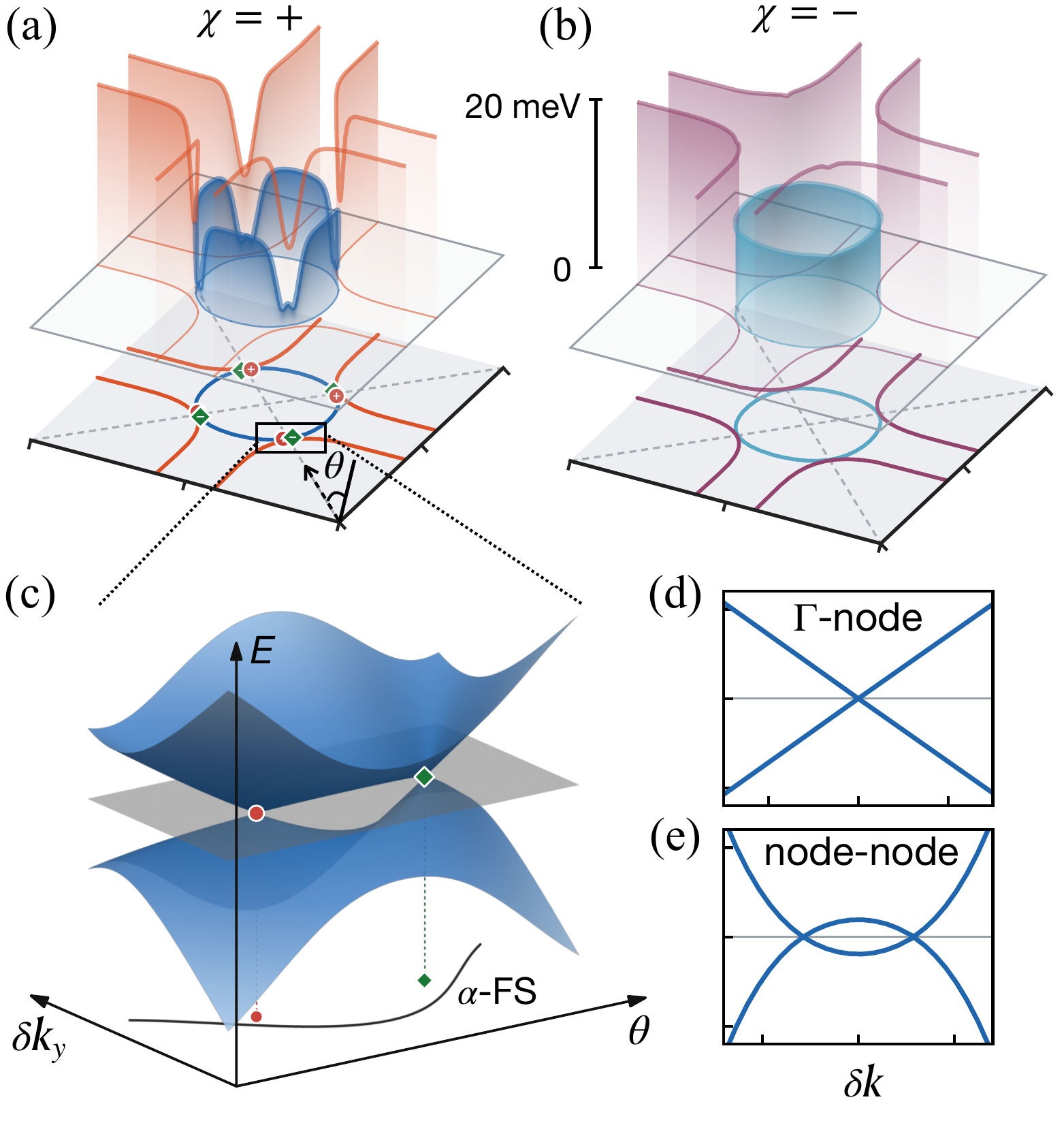}
 \caption{\label{fignode} (a),(b) BdG gap on the normal-state Fermi surfaces of helicity $\chi=+$ and $\chi=-$.  The lower planes show the Fermi surfaces, and in (a) the nodes with chiral charges $Q=+1$ (red circles) and $-1$ (green diamonds), the angle $\theta$ at M counted clockwise from $\mathrm{M}$-$\Gamma$.
(c) The energy dispersion near the node pair, versus $\theta$ and the radial offset from the nodes.  The base plane shows the $\chi=+$ $\alpha$ Fermi surface and the node projections.
(d),(e) BdG bands along the line from $\Gamma$ through a node and along the line through both nodes.}
\end{figure}

From Eqs.~\eqref{eq:straddle} and~\eqref{eq:straddle1}, it can be inferred that the following two conditions are necessary for the topological superconductivity, $V \cdot \lambda$ being nonzero and the pairing orders satisfying $|\Delta_\perp|>|\Delta_\parallel|$. In $A_{\rm c}^2$ in Eq.~\eqref{eq:straddle1}, the terms in the first bracket account for the intraband pairing which contribute sign changes of the pairing orders on the Fermi surfaces in the topological phase transition~\cite{QiHughesZhang2010}, while the terms in the second bracket correspond to the interband pairing which modifies the phase boundary. This can be clearly shown by projecting the superconducting pairing onto the band basis. For instance, along $\Gamma$-$\mathrm{M}$ in the $M_{x\bar{y}}$ subspace with eigenvalue $\ii \zeta$, the pairing on the Fermi surfaces takes the form (SM)
\begin{equation}
	\Delta_{n\zeta}=\Delta_\parallel+n\Delta_\perp
	\frac{t_\perp}{\sqrt{t_\perp^2+ (V + \zeta r)^2}},
	\label{eq:projected}
\end{equation}
where $n = \pm$ stands for the two eigenstates for $h_\zeta$ in Eq.~\eqref{eq:mirror-bdg}, and $t_\perp$ and $r$ take the values at the Fermi points along $\Gamma$-$\mathrm{M}$. In the weak-pairing condition, the above topological mirror winding number can be calculated according to the Fermi-surface formula $w_\zeta=\frac12\sum_j\operatorname{sgn}\bigl[v_\zeta(k_j)\,\Delta_\zeta(k_j)\bigr]$, where $j$ runs over the Fermi points in mirror sector $\ii \zeta$ and $v_\zeta$ ($\Delta_\zeta$) is the Fermi velocity (superconducting pairing) on the Fermi point~\cite{QiHughesZhang2010,ZhangFermi2026}. The Fermi-surface formula indicates that the winding number changes when the pairing signs on the Fermi points change, and the phase transition occurs at $\Delta_{n\zeta} = 0$ in Eq.~\eqref{eq:projected} which gives exactly the results in Eq.~\eqref{eq:straddle} by omitting the interband pairing contributions in $A_{\rm c}$. Moreover, it can be deduced from Eq.~\eqref{eq:projected} that (i) a large $| V + \zeta r |$ is conducive to the topological phase transition, and (ii) the topological phase transition tends to (cannot) occur on Fermi points with $n = +$ ($n = -$), since it satisfies $t_\perp > 0$, $| \Delta_\perp | > | \Delta_\parallel |$ and $\Delta_\perp \Delta_\parallel < 0$ in La$_3$Ni$_2$O$_7$. The superconducting gap on the Fermi surfaces shown in Fig.~\ref{fignode}(a),(b) directly corroborates the conclusions, where the gap is strongly suppressed along $\Gamma$-$\mathrm{M}$ on the Fermi surfaces in Fig.~\ref{fignode}(a) while it is slightly modified in Fig.~\ref{fignode}(b).

\runinsec{Orbital hybridization and bulk Dirac nodes}
In the above, we have focused on $\Gamma$-$\mathrm{M}$. Away from the line, topological Dirac nodes develop accompanying the emergence of the topological mirror superconductivity on $\Gamma$-$\mathrm{M}$. In the topological phase diagram in Fig.~\ref{figwindow}(a), when it moves from the trivial phase towards the nontrivial phase, the system first becomes gapless on $\Gamma$-$\mathrm{M}$ at the phase boundary. Further into the nontrivial phase, superconducting gap reopens on $\Gamma$-$\mathrm{M}$, with the nodal points on $\Gamma$-$\mathrm{M}$ splitting and evolving into the Dirac nodes in Fig.~\ref{figwindow}(c). The eight Dirac nodes are related by the $C_{4v}$ point group symmetry, and each Dirac node carries a topological charge~\cite{Beri2010,Matsuura2013}
\begin{equation}
	Q_j=-\frac{1}{4\pi\ii}\oint_{C_j}\mathrm{d}\kk\cdot
	\operatorname{Tr}\!\left[\mathcal S H_{\rm BdG}^{-1}(\kk)
	\nabla_{\kk}H_{\rm BdG}(\kk)\right],
	\label{eq:node-charge}
\end{equation}
protected by the chiral symmetry. Here, $C_j$ is a counterclockwise loop enclosing the $j$-th node and $\mathcal S$ is the chiral symmetry operator. The Dirac nodes related by rotations have the same charge, while those related by mirror reflections carry opposite charges. We label the topological charge of the Dirac nodes with different colors in Fig.~\ref{figwindow}(c).

Physically, the Dirac nodes arise from the combined effects of the orbital hybridization, the layer-asymmetric potential and the staggered Rashba SOC. It is intuitive to estimate the pairing on the Fermi surfaces based on Eq.~\eqref{eq:projected}. Though obtained on $\Gamma$-$\mathrm{M}$, Eq.~\eqref{eq:projected} can be used near the line based on perturbation theory. Since the Fermi surfaces are dominated by the $d_{x^2-y^2}$ orbital as shown in Fig.~\ref{figDFT}(d), $t_\perp$ corresponds to the hopping between the $d_{x^2-y^2}$ orbitals in different layers. Off $\Gamma$-$\mathrm{M}$ the $d_{x^2-y^2}$ orbital hybridizes with the $d_{z^2}$ orbital, and the hybridization strength grows as it moves further away from the line. Such orbital hybridization makes additional contribution to $t_\perp$ through high-order processes, meaning larger $t_\perp$ for Fermi point further away from $\Gamma$-$\mathrm{M}$. Recalling that the topological superconductivity on $\Gamma$-$\mathrm{M}$ arises due to the dominating $| V + \zeta r |$ in Eq.~\eqref{eq:projected}, on the Fermi surface contributing to the topological phase transition we must have $\Delta_{n \zeta} = 0$ at some $\kk$, since $t_\perp$ gradually becomes dominant as $\kk$ moves away from $\Gamma$-$\mathrm{M}$, leading to Dirac nodes. With the parameters fit for La$_3$Ni$_2$O$_7$, the Dirac nodes are actually off the Fermi surfaces slightly due to its finite interband pairing, leaving a deep gap minima about $1.8$~meV on the Fermi surfaces, as shown in Figs.~\ref{fignode}(a)(c).

In accordance with the Dirac nodes, zero-energy flat band exists on the edge~\cite{SatoFlat2011}. To be specific, we take the [21] edge for instance and present the numerical results in Fig.~\ref{figwindow}(d). The zero-energy flat band in Fig.~\ref{figwindow}(d) can be understood as follows. In Fig.~\ref{figwindow}(c), on each straight line with fixed $k_{21}$ in the Brillouin zone the winding number changes by $Q$, when the line goes across the Dirac node with topological charge $Q$ as $k_{21}$ varies. For the node distribution in Fig.~\ref{figwindow}(c), it must result in $k_{21}$ segments where the line carries nontrivial winding number as sketched in Fig.~\ref{figwindow}(c). Correspondingly, zero-energy flat band appears on the [21] edge, as presented in Fig.~\ref{figwindow}(d). However, on the [11] edge no flat band exists as the Dirac nodes with opposite charges always project onto the same edge momentum. And the Majorana Kramers pair splits immediately as $k_{11}$ deviates from zero as presented in Fig.~\ref{figwindow}(b).

\runinsec{Discussion}
Our work highlights the exotic interplay between pairing, interlayer hopping, and SOC in the $d_{x^2-y^2}$ orbital of bilayer nickelates, where these energy scales are comparable.
In particular, the topological window is set by $t_\perp(k_0)$, $\Delta_\perp$, and $\Delta_\parallel$ through $A_{\rm c}$ in Eq.~\eqref{eq:straddle1}, and all three are of order $10$~meV.
The effective interlayer hopping for the $d_{x^2-y^2}$ orbital is weak, as reflected in the nearly touching $\alpha$ and $\beta$ Fermi pockets along $\Gamma$-$\mathrm{M}$ observed by ARPES~\cite{YangARPES2024}.
Since these two pockets carry pairing of opposite signs, the pairing reverses its sign along the diagonal within an unusually small momentum range.
Realizing the topological superconductivity therefore does not require strong layer asymmetry or SOC, and strengths comparable to the effective interlayer hopping can suffice, making this regime experimentally accessible according to our first-principles calculations.
These features make bilayer nickelate films an exceptionally promising platform for intrinsic topological superconductivity at high temperatures.
Superconducting films grown under compressive strain on different substrates~\cite{Ko2025,Zhou2025,XHChen2026} offer opportunities to tune these scales through substrate selection and strain. The layer-asymmetric potential also varies with the distance of NiO$_2$ bilayers from the interface, potentially favoring the topological phase near the interface. The neglected layer-uniform Rashba term $h'_{\rm SOC}$ shifts the helicity branches without changing their layer composition and only weakly affects the topological window in the parameter range considered (SM). The topological phase also persists in the presence of the bonding $\gamma$ pocket.

\begin{figure}[b]
 \centering
 \includegraphics[width=0.95\columnwidth]{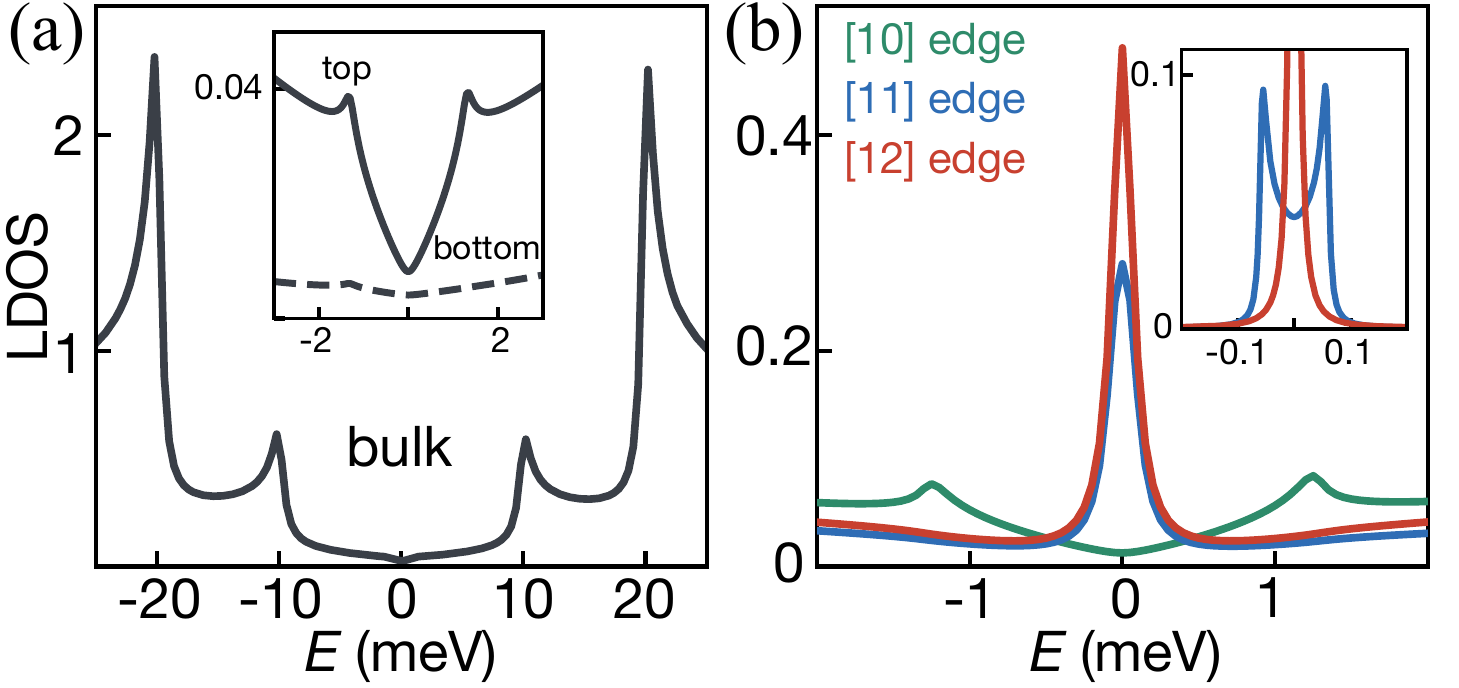}
 \caption{\label{figLDOS}Electron local density of states (LDOS) summed over the two layers.  (a) Bulk, with a broadening of $0.3$~meV.  The inset shows the top (solid) and bottom (dashed) layers separately.  (b) Surface LDOS at the [10], [11], and [21] edges, with a broadening of $0.08$~meV.  The inset shows the [11] and [21] edges with a broadening of $0.005$~meV.}
\end{figure}

The tunable bulk nodes associated with the topological transition provide a possible probe of dominant interlayer pairing. As the layer-asymmetric potential is tuned into the topological window at finite SOC, the superconducting state evolves from fully gapped and trivial to nodal and topological. In thin films, NiO$_2$ bilayers farther from and closer to the interface may therefore exhibit U-shaped and V-shaped scanning tunneling microscopy (STM) spectra, respectively. Interestingly, in the topological regime, the V-shaped bulk tunneling spectrum and zero-energy peaks at the [11] and [21] edges, but not at the [10] edge, shown in Fig.~\ref{figLDOS}(a),(b), resemble signatures of $d$-wave pairing. A distinguishing feature, however, is the layer polarization of the nodal quasiparticles: their spectral weight resides predominantly on one NiO$_2$ layer, as shown in the inset of Fig.~\ref{figLDOS}(a). An external electric field that reverses the layer-asymmetric potential can transfer their spectral weight to the opposite layer, changing the top-layer spectrum from V-shaped to nearly fully gapped. By contrast, the symmetry-enforced nodes of $d_{x^2-y^2}$ pairing remain on the diagonals under such a potential. This electrically tunable layer dependence would provide strong evidence for dominant interlayer pairing.

\runinsec{Note added}
Near the end of the paper, we became aware of a work~\cite{Scholten2026}, which studied nodes arising from orbital-anti-phase pairing in a mirror-symmetric spinless bilayer nickelate model.
In our case, we obtain nonzero mirror winding number and a Majorana Kramers pair at the [11] edge.

\begin{acknowledgments}
Z.Z. and H.C.P. acknowledge support from the Future Science Awards Foundation and the Hong Kong Research Grants Council (Grant Nos.~16310625 and AoE/P-604/25R).
C.L. is supported by the start-up fund from Hefei National Laboratory.
X.W. is supported by the National Key R\&D Program of China (Grant No.~2023YFA1407300) and the National Natural Science Foundation of China (Grant Nos.~12574151, 12447103, and 12447101).
J.H. is supported by the National Natural Science Foundation of China (Grant No.~12494594) and the New Cornerstone Investigator Program.
S.Q. is supported by the National Natural Science Foundation of China (Grant No.~12304163) and the Beijing Institute of Technology Research Fund Program for Young Scholars.
\end{acknowledgments}

\bibliography{references}

@article{Sun2023,
  author = {Sun, Hualei and Huo, Mengwu and Hu, Xunwu and Li, Jingyuan and Liu, Zengjia and Han, Yifeng and Tang, Lingyun and Mao, Zhongquan and Yang, Pengtao and Wang, Bosen and Cheng, Jinguang and Yao, Dao-Xin and Zhang, Guang-Ming and Wang, Meng},
  title = {{Signatures of superconductivity near 80 K in a nickelate under high pressure}},
  journal = {Nature},
  volume = {621},
  pages = {493--498},
  year = {2023},
  doi = {10.1038/s41586-023-06408-7}
}

@article{Ko2025,
  author = {Ko, Eun Kyo and Yu, Yijun and Liu, Yidi and Bhatt, Lopa and Li, Jiarui and Thampy, Vivek and Kuo, Cheng-Tai and Wang, Bai Yang and Lee, Yonghun and Lee, Kyuho and Lee, Jun-Sik and Goodge, Berit H. and Muller, David A. and Hwang, Harold Y.},
  title = {{Signatures of ambient pressure superconductivity in thin film La$_3$Ni$_2$O$_7$}},
  journal = {Nature},
  volume = {638},
  pages = {935--940},
  year = {2025},
  doi = {10.1038/s41586-024-08525-3}
}

@article{Zhou2025,
  author = {Zhou, Guangdi and Lv, Wei and Wang, Heng and Nie, Zihao and Chen, Yaqi and Li, Yueying and Huang, Haoliang and Chen, Wei-Qiang and Sun, Yu-Jie and Xue, Qi-Kun and Chen, Zhuoyu},
  title = {{Ambient-pressure superconductivity onset above 40 K in (La,Pr)$_3$Ni$_2$O$_7$ films}},
  journal = {Nature},
  volume = {640},
  pages = {641--646},
  year = {2025},
  doi = {10.1038/s41586-025-08755-z}
}

@article{Le2025,
  author = {Le, Congcong and Zhan, Jun and Wu, Xianxin and Hu, Jiangping},
  title = {{Opposite-Mirror-Parity Scattering as the Origin of Superconductivity in Strained Bilayer Nickelates}},
  journal = {arXiv:2501.14665},
  year = {2025},
  eprint = {2501.14665},
  archivePrefix = {arXiv},
  primaryClass = {cond-mat.supr-con}
}

@article{Gaidamauskas2014,
  author = {Gaidamauskas, E. and Paaske, J. and Flensberg, K.},
  title = {{Majorana Bound States in Two-Channel Time-Reversal-Symmetric Nanowire Systems}},
  journal = {Phys. Rev. Lett.},
  volume = {112},
  pages = {126402},
  year = {2014},
  doi = {10.1103/PhysRevLett.112.126402}
}

@article{ZhangKaneMele2013,
  author = {Zhang, Fan and Kane, C. L. and Mele, E. J.},
  title = {{Time-Reversal-Invariant Topological Superconductivity and Majorana Kramers Pairs}},
  journal = {Phys. Rev. Lett.},
  volume = {111},
  pages = {056402},
  year = {2013},
  doi = {10.1103/PhysRevLett.111.056402}
}

@article{ZhangDasSarma2021,
  author = {Zhang, Rui-Xing and Das Sarma, S.},
  title = {{Intrinsic Time-Reversal-Invariant Topological Superconductivity in Thin Films of Iron-Based Superconductors}},
  journal = {Phys. Rev. Lett.},
  volume = {126},
  pages = {137001},
  year = {2021},
  doi = {10.1103/PhysRevLett.126.137001}
}

@article{WangSTM2026,
  author = {Wang, Xinxin and Chen, Yaqi and Ding, Cui and Xu, Lizhi and Miao, Jian-Jian and Zhou, Guangdi and Chen, Zhuoyu and Sun, Yu-Jie and Jia, Jin-Feng and Xue, Qi-Kun},
  title = {{Atomically resolved intrinsic superconducting gap in (La,Pr)$_3$Ni$_2$O$_7$ films}},
  journal = {arXiv:2605.14806},
  year = {2026},
  eprint = {2605.14806},
  archivePrefix = {arXiv},
  primaryClass = {cond-mat.supr-con}
}

@article{Qin2022,
  author = {Qin, Shengshan and Fang, Chen and Zhang, Fu-Chun and Hu, Jiangping},
  title = {{Topological Superconductivity in an Extended $s$-Wave Superconductor and Its Implication to Iron-Based Superconductors}},
  journal = {Phys. Rev. X},
  volume = {12},
  pages = {011030},
  year = {2022},
  doi = {10.1103/PhysRevX.12.011030}
}

@article{Zhang2024,
  author = {Zhang, Zhongyi and Wu, Zhenfei and Fang, Chen and Zhang, Fu-Chun and Hu, Jiangping and Wang, Yuxuan and Qin, Shengshan},
  title = {{Topological superconductivity from unconventional band degeneracy with conventional pairing}},
  journal = {Nat. Commun.},
  volume = {15},
  pages = {7971},
  year = {2024},
  doi = {10.1038/s41467-024-52156-1}
}

@article{SatoAndo2017,
  author = {Sato, Masatoshi and Ando, Yoichi},
  title = {{Topological superconductors: a review}},
  journal = {Rep. Prog. Phys.},
  volume = {80},
  pages = {076501},
  year = {2017},
  doi = {10.1088/1361-6633/aa6ac7}
}

@article{QiHughesZhang2010,
  author = {Qi, Xiao-Liang and Hughes, Taylor L. and Zhang, Shou-Cheng},
  title = {{Topological invariants for the {Fermi} surface of a time-reversal-invariant superconductor}},
  journal = {Phys. Rev. B},
  volume = {81},
  pages = {134508},
  year = {2010},
  doi = {10.1103/PhysRevB.81.134508}
}

@article{ZhangQPI2026,
  author = {Zhang, Zhongyi and Zhan, Jun and Le, Congcong and Po, Hoi Chun and Hu, Jiangping and Wu, Xianxin},
  title = {{Mirror-selective quasiparticle interference in bilayer nickelate superconductor}},
  journal = {npj Quantum Mater.},
  year = {2026},
  doi = {10.1038/s41535-026-00913-4}
}

@article{ZhangZunger2014,
  author = {Zhang, Xiuwen and Liu, Qihang and Luo, Jun-Wei and Freeman, Arthur J. and Zunger, Alex},
  title = {{Hidden spin polarization in inversion-symmetric bulk crystals}},
  journal = {Nat. Phys.},
  volume = {10},
  pages = {387--393},
  year = {2014},
  doi = {10.1038/nphys2933}
}

@article{FischerReview2023,
  author = {Fischer, Mark H. and Sigrist, Manfred and Agterberg, Daniel F. and Yanase, Youichi},
  title = {{Superconductivity and Local Inversion-Symmetry Breaking}},
  journal = {Annu. Rev. Condens. Matter Phys.},
  volume = {14},
  pages = {153-172},
  year = {2023},
  doi = {10.1146/annurev-conmatphys-040521-042511}
}

@article{WuBiS2_2017,
  author = {Wu, Shi-Long and Sumida, Kazuki and Miyamoto, Koji and Taguchi, Kazuaki and Yoshikawa, Tomoki and Kimura, Akio and Ueda, Yoshifumi and Arita, Masashi and Nagao, Masanori and Watauchi, Satoshi and Tanaka, Isao and Okuda, Taichi},
  title = {{Direct evidence of hidden local spin polarization in a centrosymmetric superconductor {LaO}$_{0.55}${F}$_{0.45}${BiS}$_2$}},
  journal = {Nat. Commun.},
  volume = {8},
  pages = {1919},
  year = {2017},
  doi = {10.1038/s41467-017-02058-2}
}

@article{YuanZunger2019,
  author = {Yuan, Linding and Liu, Qihang and Zhang, Xiuwen and Luo, Jun-Wei and Li, Shu-Shen and Zunger, Alex},
  title = {{Uncovering and tailoring hidden Rashba spin--orbit splitting in centrosymmetric crystals}},
  journal = {Nat. Commun.},
  volume = {10},
  pages = {906},
  year = {2019},
  doi = {10.1038/s41467-019-08836-4}
}

@article{AgterbergFeSe2017,
  author = {Agterberg, D. F. and Shishidou, T. and O'Halloran, J. and Brydon, P. M. R. and Weinert, M.},
  title = {{Resilient Nodeless $d$-Wave Superconductivity in Monolayer {FeSe}}},
  journal = {Phys. Rev. Lett.},
  volume = {119},
  pages = {267001},
  year = {2017},
  doi = {10.1103/PhysRevLett.119.267001}
}

@article{ZhangLiuHSP2020,
  author = {Zhang, Yingjie and Liu, Pengfei and Sun, Hongyi and Zhao, Shixuan and Xu, Hu and Liu, Qihang},
  title = {{Symmetry-Assisted Protection and Compensation of Hidden Spin Polarization in Centrosymmetric Systems}},
  journal = {Chin. Phys. Lett.},
  volume = {37},
  pages = {087105},
  year = {2020},
  doi = {10.1088/0256-307X/37/8/087105}
}

@article{Luo2023,
  author = {Luo, Zhihui and Hu, Xunwu and Wang, Meng and Wú, Wéi and Yao, Dao-Xin},
  title = {{Bilayer Two-Orbital Model of {La}$_3${Ni}$_2${O}$_7$ under Pressure}},
  journal = {Phys. Rev. Lett.},
  volume = {131},
  pages = {126001},
  year = {2023},
  doi = {10.1103/PhysRevLett.131.126001}
}

@article{LiuApical2023,
  author = {Liu, Yu-Bo and Mei, Jia-Wei and Ye, Fei and Chen, Wei-Qiang and Yang, Fan},
  title = {{$s_\pm$-Wave Pairing and the Destructive Role of Apical-Oxygen Deficiencies in {La}$_3${Ni}$_2${O}$_7$ under Pressure}},
  journal = {Phys. Rev. Lett.},
  volume = {131},
  pages = {236002},
  year = {2023},
  doi = {10.1103/PhysRevLett.131.236002}
}

@article{Qu2024,
  author = {Qu, Xing-Zhou and Qu, Dai-Wei and Chen, Jialin and Wu, Congjun and Yang, Fan and Li, Wei and Su, Gang},
  title = {{Bilayer $t$-$J$-$J_\perp$ Model and Magnetically Mediated Pairing in the Pressurized Nickelate {La}$_3${Ni}$_2${O}$_7$}},
  journal = {Phys. Rev. Lett.},
  volume = {132},
  pages = {036502},
  year = {2024},
  doi = {10.1103/PhysRevLett.132.036502}
}

@article{QiHughesRaghuZhang2009,
  author = {Qi, Xiao-Liang and Hughes, Taylor L. and Raghu, S. and Zhang, Shou-Cheng},
  title = {{Time-Reversal-Invariant Topological Superconductors and Superfluids in Two and Three Dimensions}},
  journal = {Phys. Rev. Lett.},
  volume = {102},
  pages = {187001},
  year = {2009},
  doi = {10.1103/PhysRevLett.102.187001}
}

@article{SatoFlat2011,
  author = {Sato, Masatoshi and Tanaka, Yukio and Yada, Keiji and Yokoyama, Takehito},
  title = {{Topology of {Andreev} bound states with flat dispersion}},
  journal = {Phys. Rev. B},
  volume = {83},
  pages = {224511},
  year = {2011},
  doi = {10.1103/PhysRevB.83.224511}
}

@article{Hu1994,
  author = {Hu, Chia-Ren},
  title = {{Midgap surface states as a novel signature for {$d_{x_a^2-x_b^2}$}-wave superconductivity}},
  journal = {Phys. Rev. Lett.},
  volume = {72},
  pages = {1526},
  year = {1994},
  doi = {10.1103/PhysRevLett.72.1526}
}

@article{PotterLee2014,
  author = {Potter, Andrew C. and Lee, Patrick A.},
  title = {{Edge Ferromagnetism from {Majorana} Flat Bands: Application to Split Tunneling-Conductance Peaks in High-{$T_c$} Cuprate Superconductors}},
  journal = {Phys. Rev. Lett.},
  volume = {112},
  pages = {117002},
  year = {2014},
  doi = {10.1103/PhysRevLett.112.117002}
}

@article{Beri2010,
  author = {Béri, B.},
  title = {{Topologically stable gapless phases of time-reversal-invariant superconductors}},
  journal = {Phys. Rev. B},
  volume = {81},
  pages = {134515},
  year = {2010},
  doi = {10.1103/PhysRevB.81.134515}
}

@article{Matsuura2013,
  author = {Matsuura, Shunji and Chang, Po-Yao and Schnyder, Andreas P and Ryu, Shinsei},
  title = {{Protected boundary states in gapless topological phases}},
  journal = {New J. Phys.},
  volume = {15},
  pages = {065001},
  year = {2013},
  doi = {10.1088/1367-2630/15/6/065001}
}

@article{ChiuRMP2016,
  author = {Chiu, Ching-Kai and Teo, Jeffrey C. Y. and Schnyder, Andreas P. and Ryu, Shinsei},
  title = {{Classification of topological quantum matter with symmetries}},
  journal = {Rev. Mod. Phys.},
  volume = {88},
  pages = {035005},
  year = {2016},
  doi = {10.1103/RevModPhys.88.035005}
}

@article{AltlandZirnbauer1997,
  author = {Altland, Alexander and Zirnbauer, Martin R.},
  title = {{Nonstandard symmetry classes in mesoscopic normal-superconducting hybrid structures}},
  journal = {Phys. Rev. B},
  volume = {55},
  pages = {1142-1161},
  year = {1997},
  doi = {10.1103/PhysRevB.55.1142}
}

@article{ZhangZeroRes2024,
  author = {Zhang, Yanan and Su, Dajun and Huang, Yanen and Shan, Zhaoyang and Sun, Hualei and Huo, Mengwu and Ye, Kaixin and Zhang, Jiawen and Yang, Zihan and Xu, Yongkang and Su, Yi and Li, Rui and Smidman, Michael and Wang, Meng and Jiao, Lin and Yuan, Huiqiu},
  title = {{High-temperature superconductivity with zero resistance and strange-metal behaviour in {La}$_3${Ni}$_2${O}$_{7-\delta}$}},
  journal = {Nat. Phys.},
  volume = {20},
  pages = {1269-1273},
  year = {2024},
  doi = {10.1038/s41567-024-02515-y}
}

@article{WangBulk2024,
  author = {Wang, Ningning and Wang, Gang and Shen, Xiaoling and Hou, Jun and Luo, Jun and Ma, Xiaoping and Yang, Huaixin and Shi, Lifen and Dou, Jie and Feng, Jie and Yang, Jie and Shi, Yunqing and Ren, Zhian and Ma, Hanming and Yang, Pengtao and Liu, Ziyi and Liu, Yue and Zhang, Hua and Dong, Xiaoli and Wang, Yuxin and Jiang, Kun and Hu, Jiangping and Nagasaki, Shoko and Kitagawa, Kentaro and Calder, Stuart and Yan, Jiaqiang and Sun, Jianping and Wang, Bosen and Zhou, Rui and Uwatoko, Yoshiya and Cheng, Jinguang},
  title = {{Bulk high-temperature superconductivity in pressurized tetragonal La$_2$PrNi$_2$O$_7$}},
  journal = {Nature},
  volume = {634},
  pages = {579-584},
  year = {2024},
  doi = {10.1038/s41586-024-07996-8}
}

@article{Zhou2026,
  author = {Zhou, Guangdi and Wang, Heng and Huang, Haoliang and Chen, Yaqi and Peng, Fei and Lv, Wei and Nie, Zihao and Wang, Wei and Jia, Jin-Feng and Xue, Qi-Kun and Chen, Zhuoyu},
  title = {{Superconductivity onset above 60 {K} in ambient-pressure nickelate films}},
  journal = {Natl. Sci. Rev.},
  volume = {13},
  pages = {nwag151},
  year = {2026},
  doi = {10.1093/nsr/nwag151}
}

@article{Lu2024,
  author = {Lu, Chen and Pan, Zhiming and Yang, Fan and Wu, Congjun},
  title = {{Interlayer-Coupling-Driven High-Temperature Superconductivity in {La}$_3${Ni}$_2${O}$_7$ under Pressure}},
  journal = {Phys. Rev. Lett.},
  volume = {132},
  pages = {146002},
  year = {2024},
  doi = {10.1103/PhysRevLett.132.146002}
}

@article{YangSpm2023,
  author = {Yang, Qing-Geng and Wang, Da and Wang, Qiang-Hua},
  title = {{Possible $s_\pm$-wave superconductivity in {La}$_3${Ni}$_2${O}$_7$}},
  journal = {Phys. Rev. B},
  volume = {108},
  pages = {L140505},
  year = {2023},
  doi = {10.1103/PhysRevB.108.L140505}
}

@article{Sakakibara2024,
  author = {Sakakibara, Hirofumi and Kitamine, Naoya and Ochi, Masayuki and Kuroki, Kazuhiko},
  title = {{Possible High $T_c$ Superconductivity in {La}$_3${Ni}$_2${O}$_7$ under High Pressure through Manifestation of a Nearly Half-Filled Bilayer {Hubbard} Model}},
  journal = {Phys. Rev. Lett.},
  volume = {132},
  pages = {106002},
  year = {2024},
  doi = {10.1103/PhysRevLett.132.106002}
}

@article{ZhangDagotto2024,
  author = {Zhang, Yang and Lin, Ling-Fang and Moreo, Adriana and Maier, Thomas A. and Dagotto, Elbio},
  title = {{Structural phase transition, $s_\pm$-wave pairing, and magnetic stripe order in bilayered superconductor {La}$_3${Ni}$_2${O}$_7$ under pressure}},
  journal = {Nat. Commun.},
  volume = {15},
  pages = {2470},
  year = {2024},
  doi = {10.1038/s41467-024-46622-z}
}

@article{Gu2025,
  author = {Gu, Yuhao and Le, Congcong and Yang, Zhesen and Wu, Xianxin and Hu, Jiangping},
  title = {{Effective model and pairing tendency in the bilayer Ni-based superconductor {La}$_3${Ni}$_2${O}$_7$}},
  journal = {Phys. Rev. B},
  volume = {111},
  pages = {174506},
  year = {2025},
  doi = {10.1103/PhysRevB.111.174506}
}

@article{Christiansson2023,
  author = {Christiansson, Viktor and Petocchi, Francesco and Werner, Philipp},
  title = {{Correlated Electronic Structure of {La}$_3${Ni}$_2${O}$_7$ under Pressure}},
  journal = {Phys. Rev. Lett.},
  volume = {131},
  pages = {206501},
  year = {2023},
  doi = {10.1103/PhysRevLett.131.206501}
}

@article{Lechermann2023,
  author = {Lechermann, Frank and Gondolf, Jannik and Bötzel, Steffen and Eremin, Ilya M.},
  title = {{Electronic correlations and superconducting instability in {La}$_3${Ni}$_2${O}$_7$ under high pressure}},
  journal = {Phys. Rev. B},
  volume = {108},
  pages = {L201121},
  year = {2023},
  doi = {10.1103/PhysRevB.108.L201121}
}

@article{FuBerg2010,
  author = {Fu, Liang and Berg, Erez},
  title = {{Odd-Parity Topological Superconductors: Theory and Application to {Cu}$_x${Bi}$_2${Se}$_3$}},
  journal = {Phys. Rev. Lett.},
  volume = {105},
  pages = {097001},
  year = {2010},
  doi = {10.1103/PhysRevLett.105.097001}
}

@article{ZhangFermi2026,
  author = {Zhang, Zhongyi and Shiozaki, Ken and Fang, Chen and Ono, Seishiro},
  title = {{Fermi-surface diagnosis for topological superconductivity with $s$-wave-like pairing symmetries}},
  journal = {Nat. Commun.},
  volume = {17},
  pages = {4413},
  year = {2026},
  doi = {10.1038/s41467-026-72811-z}
}

@article{Haim2016NoGo,
  author = {Haim, Arbel and Berg, Erez and Flensberg, Karsten and Oreg, Yuval},
  title = {{No-go theorem for a time-reversal invariant topological phase in noninteracting systems coupled to conventional superconductors}},
  journal = {Phys. Rev. B},
  volume = {94},
  pages = {161110(R)},
  year = {2016},
  doi = {10.1103/PhysRevB.94.161110}
}

@article{Deng2012TRISwave,
  author = {Deng, Shusa and Viola, Lorenza and Ortiz, Gerardo},
  title = {{{Majorana} Modes in Time-Reversal Invariant {$s$}-Wave Topological Superconductors}},
  journal = {Phys. Rev. Lett.},
  volume = {108},
  pages = {036803},
  year = {2012},
  doi = {10.1103/PhysRevLett.108.036803}
}

@article{Ono2024Classification,
  author = {Ono, Seishiro and Shiozaki, Ken and Watanabe, Haruki},
  title = {{Classification of time-reversal symmetric topological superconducting phases for conventional pairing symmetries}},
  journal = {Phys. Rev. B},
  volume = {109},
  pages = {214502},
  year = {2024},
  doi = {10.1103/PhysRevB.109.214502}
}

@article{Scholten2026,
  author = {Scholten, Marius and B{\"o}tzel, Steffen and Lechermann, Frank and Fernandes, Rafael M. and Eremin, Ilya M.},
  title = {{Nodal Orbital-Anti-Phase Superconducting State in Bilayer Nickelates}},
  journal = {arXiv:2609.29263},
  year = {2026},
  eprint = {2609.29263},
  archivePrefix = {arXiv},
  primaryClass = {cond-mat.supr-con}
}

@article{ZhangLin2023,
  author = {Zhang, Yang and Lin, Ling-Fang and Moreo, Adriana and Dagotto, Elbio},
  title = {{Electronic structure, dimer physics, orbital-selective behavior, and magnetic tendencies in the bilayer nickelate superconductor {La}$_3${Ni}$_2${O}$_7$ under pressure}},
  journal = {Phys. Rev. B},
  volume = {108},
  pages = {L180510},
  year = {2023},
  doi = {10.1103/PhysRevB.108.L180510}
}

@article{LiuOptical2024,
  author = {Liu, Zhe and Huo, Mengwu and Li, Jie and Li, Qing and Liu, Yuecong and Dai, Yaomin and Zhou, Xiaoxiang and Hao, Jiahao and Lu, Yi and Wang, Meng and Wen, Hai-Hu},
  title = {{Electronic correlations and partial gap in the bilayer nickelate {La}$_3${Ni}$_2${O}$_7$}},
  journal = {Nat. Commun.},
  volume = {15},
  pages = {7570},
  year = {2024},
  doi = {10.1038/s41467-024-52001-5}
}

@article{YangARPES2024,
  author = {Yang, Jiangang and Sun, Hualei and Hu, Xunwu and Xie, Yuyang and Miao, Taimin and Luo, Hailan and Chen, Hao and Liang, Bo and Zhu, Wenpei and Qu, Gexing and Chen, Cui-Qun and Huo, Mengwu and Huang, Yaobo and Zhang, Shenjin and Zhang, Fengfeng and Yang, Feng and Wang, Zhimin and Peng, Qinjun and Mao, Hanqing and Liu, Guodong and Xu, Zuyan and Qian, Tian and Yao, Dao-Xin and Wang, Meng and Zhao, Lin and Zhou, X. J.},
  title = {{Orbital-dependent electron correlation in double-layer nickelate {La}$_3${Ni}$_2${O}$_7$}},
  journal = {Nat. Commun.},
  volume = {15},
  pages = {4373},
  year = {2024},
  doi = {10.1038/s41467-024-48701-7}
}

@article{SunARPES2026,
  author = {Sun, Wenjie and Jiang, Zhicheng and Hao, Bo and Yan, Shengjun and Zhang, Hongyi and Wang, Maosen and Yang, Yang and Sun, Haoying and Liu, Zhengtai and Ji, Dianxiang and Gu, Zhengbin and Zhou, Jian and Shen, Dawei and Feng, Donglai and Nie, Yuefeng},
  title = {{Observation of superconductivity-induced leading-edge gap in a bilayer nickelate}},
  journal = {Nat. Phys.},
  volume = {22},
  pages = {1409--1415},
  year = {2026},
  doi = {10.1038/s41567-026-03396-z}
}

@article{ShenDMRG2023,
  author = {Shen, Yang and Qin, Mingpu and Zhang, Guang-Ming},
  title = {{Effective Bi-Layer Model {Hamiltonian} and Density-Matrix Renormalization Group Study for the High-$T_c$ Superconductivity in {La}$_3${Ni}$_2${O}$_7$ under High Pressure}},
  journal = {Chin. Phys. Lett.},
  volume = {40},
  pages = {127401},
  year = {2023},
  doi = {10.1088/0256-307X/40/12/127401}
}

@article{OhZhang2023,
  author = {Oh, Hanbit and Zhang, Ya-Hui},
  title = {{Type-{II} $t$-$J$ model and shared superexchange coupling from {Hund's} rule in superconducting {La}$_3${Ni}$_2${O}$_7$}},
  journal = {Phys. Rev. B},
  volume = {108},
  pages = {174511},
  year = {2023},
  doi = {10.1103/PhysRevB.108.174511}
}

@article{Liao2023,
  author = {Liao, Zhiguang and Chen, Lei and Duan, Guijing and Wang, Yiming and Liu, Changle and Yu, Rong and Si, Qimiao},
  title = {{Electron correlations and superconductivity in {La}$_3${Ni}$_2${O}$_7$ under pressure tuning}},
  journal = {Phys. Rev. B},
  volume = {108},
  pages = {214522},
  year = {2023},
  doi = {10.1103/PhysRevB.108.214522}
}

@article{YangZhangZhang2023,
  author = {Yang, Yi-feng and Zhang, Guang-Ming and Zhang, Fu-Chun},
  title = {{Interlayer valence bonds and two-component theory for high-$T_c$ superconductivity of {La}$_3${Ni}$_2${O}$_7$ under pressure}},
  journal = {Phys. Rev. B},
  volume = {108},
  pages = {L201108},
  year = {2023},
  doi = {10.1103/PhysRevB.108.L201108}
}

@article{ZhangTrends2023,
  author = {Zhang, Yang and Lin, Ling-Fang and Moreo, Adriana and Maier, Thomas A. and Dagotto, Elbio},
  title = {{Trends in electronic structures and $s_\pm$-wave pairing for the rare-earth series in bilayer nickelate superconductor $R_3${Ni}$_2${O}$_7$}},
  journal = {Phys. Rev. B},
  volume = {108},
  pages = {165141},
  year = {2023},
  doi = {10.1103/PhysRevB.108.165141}
}

@article{QinYang2023,
  author = {Qin, Qiong and Yang, Yi-feng},
  title = {{High-$T_c$ superconductivity by mobilizing local spin singlets and possible route to higher $T_c$ in pressurized {La}$_3${Ni}$_2${O}$_7$}},
  journal = {Phys. Rev. B},
  volume = {108},
  pages = {L140504},
  year = {2023},
  doi = {10.1103/PhysRevB.108.L140504}
}

@article{Tian2024,
  author = {Tian, Yi-Heng and Chen, Yin and Wang, Jia-Ming and He, Rong-Qiang and Lu, Zhong-Yi},
  title = {{Correlation effects and concomitant two-orbital $s_\pm$-wave superconductivity in {La}$_3${Ni}$_2${O}$_7$ under high pressure}},
  journal = {Phys. Rev. B},
  volume = {109},
  pages = {165154},
  year = {2024},
  doi = {10.1103/PhysRevB.109.165154}
}

@article{Luo2024,
  author = {Luo, Zhihui and Lv, Biao and Wang, Meng and W{\'u}, W{\'e}i and Yao, Dao-Xin},
  title = {{High-$T_C$ superconductivity in {La}$_3${Ni}$_2${O}$_7$ based on the bilayer two-orbital $t$-$J$ model}},
  journal = {npj Quantum Mater.},
  volume = {9},
  pages = {61},
  year = {2024},
  doi = {10.1038/s41535-024-00668-w}
}

@article{Ryee2024,
  author = {Ryee, Siheon and Witt, Niklas and Wehling, Tim O.},
  title = {{Quenched Pair Breaking by Interlayer Correlations as a Key to Superconductivity in {La}$_3${Ni}$_2${O}$_7$}},
  journal = {Phys. Rev. Lett.},
  volume = {133},
  pages = {096002},
  year = {2024},
  doi = {10.1103/PhysRevLett.133.096002}
}

@article{Xia2025,
  author = {Xia, Chengliang and Liu, Hongquan and Zhou, Shengjie and Chen, Hanghui},
  title = {{Sensitive dependence of pairing symmetry on {Ni}-$e_g$ crystal field splitting in the nickelate superconductor {La}$_3${Ni}$_2${O}$_7$}},
  journal = {Nat. Commun.},
  volume = {16},
  pages = {1054},
  year = {2025},
  doi = {10.1038/s41467-025-56206-0}
}

@article{Zhan2025,
  author = {Zhan, Jun and Gu, Yuhao and Wu, Xianxin and Hu, Jiangping},
  title = {{Cooperation between Electron-Phonon Coupling and Electronic Interaction in Bilayer Nickelates {La}$_3${Ni}$_2${O}$_7$}},
  journal = {Phys. Rev. Lett.},
  volume = {134},
  pages = {136002},
  year = {2025},
  doi = {10.1103/PhysRevLett.134.136002}
}

@article{Bejas2025,
  author = {Bejas, Mat{\'\i}as and Wu, Xianxin and Chakraborty, Debmalya and Schnyder, Andreas P. and Greco, Andr{\'e}s},
  title = {{Out-of-plane bond-order phase, superconductivity, and their competition in the $t$-$J_\parallel$-$J_\perp$ model: Possible implications for bilayer nickelates}},
  journal = {Phys. Rev. B},
  volume = {111},
  pages = {144514},
  year = {2025},
  doi = {10.1103/PhysRevB.111.144514}
}

@article{Zhan2026,
  author = {Zhan, Jun and Le, Congcong and Wu, Xianxin and Hu, Jiangping},
  title = {{Impact of nonlocal {Coulomb} repulsion on superconductivity and density-wave orders in bilayer nickelates}},
  journal = {npj Quantum Mater.},
  volume = {11},
  pages = {70},
  year = {2026},
  doi = {10.1038/s41535-026-00883-7}
}

@article{LuSMG2026,
  author = {Lu, Da-Chuan and Li, Miao and Zeng, Zhao-Yi and Hou, Wanda and Wang, Juven and Yang, Fan and You, Yi-Zhuang},
  title = {{Superconductivity from doping symmetric mass generation insulators: application to {La}$_3${Ni}$_2${O}$_7$ under pressure}},
  journal = {npj Quantum Mater.},
  year = {2026},
  doi = {10.1038/s41535-026-00925-0}
}

@article{Duan2025,
  author = {Duan, Guijing and Liao, Zhiguang and Chen, Lei and Wang, Yiming and Yu, Rong and Si, Qimiao},
  title = {{Orbital-selective correlation effects and superconducting pairing symmetry in a multiorbital $t$-$J$ model for bilayer nickelates}},
  journal = {arXiv:2502.09195},
  year = {2025},
  eprint = {2502.09195},
  archivePrefix = {arXiv},
  primaryClass = {cond-mat.supr-con}
}

@article{OhZhangPDW2025,
  author = {Oh, Hanbit and Zhang, Ya-Hui},
  title = {{Pair-density-wave superconductivity and {Anderson's} theorem in bilayer nickelates}},
  journal = {arXiv:2512.15023},
  year = {2025},
  eprint = {2512.15023},
  archivePrefix = {arXiv},
  primaryClass = {cond-mat.str-el}
}

@article{FanVishwanath2025,
  author = {Fan, Zheng-Duo and Vishwanath, Ashvin},
  title = {{Minimal two band model and experimental proposals to distinguish pairing mechanisms of the high-$T_c$ superconductor {La}$_3${Ni}$_2${O}$_7$}},
  journal = {arXiv:2512.05956},
  year = {2025},
  eprint = {2512.05956},
  archivePrefix = {arXiv},
  primaryClass = {cond-mat.str-el}
}

@article{FuKane2008,
  author = {Fu, Liang and Kane, C. L.},
  title = {{Superconducting Proximity Effect and Majorana Fermions at the Surface of a Topological Insulator}},
  journal = {Phys. Rev. Lett.},
  volume = {100},
  pages = {096407},
  year = {2008},
  doi = {10.1103/PhysRevLett.100.096407}
}

@article{Sau2010,
  author = {Sau, Jay D. and Lutchyn, Roman M. and Tewari, Sumanta and Das Sarma, S.},
  title = {{Generic New Platform for Topological Quantum Computation Using Semiconductor Heterostructures}},
  journal = {Phys. Rev. Lett.},
  volume = {104},
  pages = {040502},
  year = {2010},
  doi = {10.1103/PhysRevLett.104.040502}
}

@article{Kitaev2001,
  author = {Kitaev, A. Yu.},
  title = {{Unpaired Majorana fermions in quantum wires}},
  journal = {Phys. Usp.},
  volume = {44},
  number = {10S},
  pages = {131--136},
  year = {2001},
  doi = {10.1070/1063-7869/44/10S/S29}
}

@article{WuBoundary2020,
  author = {Wu, Xianxin and Benalcazar, Wladimir A. and Li, Yinxiang and Thomale, Ronny and Liu, Chao-Xing and Hu, Jiangping},
  title = {{Boundary-Obstructed Topological High-$T_{\rm c}$ Superconductivity in Iron Pnictides}},
  journal = {Phys. Rev. X},
  volume = {10},
  pages = {041014},
  year = {2020},
  doi = {10.1103/PhysRevX.10.041014}
}

@article{ZhangSurface2018,
  author = {Zhang, Peng and Yaji, Koichiro and Hashimoto, Takahiro and Ota, Yuichi and Kondo, Takeshi and Okazaki, Kozo and Wang, Zhijun and Wen, Jinsheng and Gu, G. D. and Ding, Hong and Shin, Shik},
  title = {{Observation of topological superconductivity on the surface of an iron-based superconductor}},
  journal = {Science},
  volume = {360},
  pages = {182--186},
  year = {2018},
  doi = {10.1126/science.aan4596}
}

@article{WangMajorana2018,
  author = {Wang, Dongfei and Kong, Lingyuan and Fan, Peng and Chen, Hui and Zhu, Shiyu and Liu, Wenyao and Cao, Lu and Sun, Yujie and Du, Shixuan and Schneeloch, John and Zhong, Ruidan and Gu, Genda and Fu, Liang and Ding, Hong and Gao, Hong-Jun},
  title = {{Evidence for Majorana bound states in an iron-based superconductor}},
  journal = {Science},
  volume = {362},
  pages = {333--335},
  year = {2018},
  doi = {10.1126/science.aao1797}
}

@article{Kong2019,
  author = {Kong, Lingyuan and Zhu, Shiyu and Papaj, Micha{\l} and Chen, Hui and Cao, Lu and Isobe, Hiroki and Xing, Yuqing and Liu, Wenyao and Wang, Dongfei and Fan, Peng and Sun, Yujie and Du, Shixuan and Schneeloch, John and Zhong, Ruidan and Gu, Genda and Fu, Liang and Gao, Hong-Jun and Ding, Hong},
  title = {{Half-integer level shift of vortex bound states in an iron-based superconductor}},
  journal = {Nat. Phys.},
  volume = {15},
  pages = {1181--1187},
  year = {2019},
  doi = {10.1038/s41567-019-0630-5}
}

@article{Nakagawa2006,
  author = {Nakagawa, Naoyuki and Hwang, Harold Y. and Muller, David A.},
  title = {{Why some interfaces cannot be sharp}},
  journal = {Nat. Mater.},
  volume = {5},
  pages = {204--209},
  year = {2006},
  doi = {10.1038/nmat1569}
}

@article{Ono2020Indicators,
  author = {Ono, Seishiro and Po, Hoi Chun and Watanabe, Haruki},
  title = {{Refined symmetry indicators for topological superconductors in all space groups}},
  journal = {Sci. Adv.},
  volume = {6},
  pages = {eaaz8367},
  year = {2020},
  doi = {10.1126/sciadv.aaz8367}
}

@article{PhysRevX.8.041056,
  author = {Liu, Qin and Chen, Chen and Zhang, Tong and Peng, Rui and Yan, Ya-Jun and Wen, Chen-Hao-Ping and Lou, Xia and Huang, Yu-Long and Tian, Jin-Peng and Dong, Xiao-Li and Wang, Guang-Wei and Bao, Wei-Cheng and Wang, Qiang-Hua and Yin, Zhi-Ping and Zhao, Zhong-Xian and Feng, Dong-Lai},
  title = {{Robust and Clean Majorana Zero Mode in the Vortex Core of High-Temperature Superconductor (Li$_{0.84}$Fe$_{0.16}$)OHFeSe}},
  journal = {Phys. Rev. X},
  volume = {8},
  pages = {041056},
  year = {2018},
  doi = {10.1103/PhysRevX.8.041056}
}

@article{Machida2019,
  author = {Machida, T. and Sun, Y. and Pyon, S. and Takeda, S. and Kohsaka, Y. and Hanaguri, T. and Sasagawa, T. and Tamegai, T.},
  title = {{Zero-energy vortex bound state in the superconducting topological surface state of Fe(Se,Te)}},
  journal = {Nat. Mater.},
  volume = {18},
  pages = {811--815},
  year = {2019},
  doi = {10.1038/s41563-019-0397-1}
}

@article{Li2022,
  author = {Li, Meng and Li, Geng and Cao, Lu and Zhou, Xingtai and Wang, Xiancheng and Jin, Changqing and Chiu, Ching-Kai and Pennycook, Stephen J. and Wang, Ziqiang and Gao, Hong-Jun},
  title = {{Ordered and tunable Majorana-zero-mode lattice in naturally strained LiFeAs}},
  journal = {Nature},
  volume = {606},
  pages = {890--895},
  year = {2022},
  doi = {10.1038/s41586-022-04744-8}
}

@article{hao2014topological,
  author = {Hao, Ningning and Hu, Jiangping},
  title = {{Topological Phases in the Single-Layer FeSe}},
  journal = {Phys. Rev. X},
  volume = {4},
  pages = {031053},
  year = {2014},
  doi = {10.1103/PhysRevX.4.031053}
}

@article{wu2016topological,
  author = {Wu, Xianxin and Qin, Shengshan and Liang, Yi and Fan, Heng and Hu, Jiangping},
  title = {{Topological characters in Fe(Te$_{1-x}$Se$_x$) thin films}},
  journal = {Phys. Rev. B},
  volume = {93},
  pages = {115129},
  year = {2016},
  doi = {10.1103/PhysRevB.93.115129}
}

@article{wang2015topological,
  author = {Wang, Zhijun and Zhang, P. and Xu, Gang and Zeng, L. K. and Miao, H. and Xu, Xiaoyan and Qian, T. and Weng, Hongming and Richard, P. and Fedorov, A. V. and Ding, H. and Dai, Xi and Fang, Zhong},
  title = {{Topological nature of the FeSe$_{0.5}$Te$_{0.5}$ superconductor}},
  journal = {Phys. Rev. B},
  volume = {92},
  pages = {115119},
  year = {2015},
  doi = {10.1103/PhysRevB.92.115119}
}

@article{xu2016topological,
  author = {Xu, Gang and Lian, Biao and Tang, Peizhe and Qi, Xiao-Liang and Zhang, Shou-Cheng},
  title = {{Topological Superconductivity on the Surface of Fe-Based Superconductors}},
  journal = {Phys. Rev. Lett.},
  volume = {117},
  pages = {047001},
  year = {2016},
  doi = {10.1103/PhysRevLett.117.047001}
}

@article{Geisler2025,
  author = {Geisler, Benjamin and Hamlin, James J. and Stewart, Gregory R. and Hennig, Richard G. and Hirschfeld, P. J.},
  title = {{Fermi surface reconstruction and enhanced spin fluctuations in strained La$_3$Ni$_2$O$_7$ on LaAlO$_3$(001) and SrTiO$_3$(001)}},
  journal = {Phys. Rev. B},
  volume = {112},
  pages = {L100506},
  year = {2025},
  doi = {10.1103/g3cz-7w7f}
}

@article{Zhao2025,
  author = {Zhao, Yi-Feng and Botana, Antia S.},
  title = {{Electronic structure of Ruddlesden-Popper nickelates: Strain to mimic the effects of pressure}},
  journal = {Phys. Rev. B},
  volume = {111},
  pages = {115154},
  year = {2025},
  doi = {10.1103/PhysRevB.111.115154}
}

@article{bhatt2026structural,
  author = {Bhatt, Lopa and Abarca Morales, Edgar and Jiang, Abigail Y. and Ko, Eun Kyo and Zhao, Yi-Feng and Schnitzer, Noah and Pan, Grace A. and Ferenc Segedin, Dan and Liu, Yidi and Yu, Yijun and Brooks, Charles M. and Botana, Antia S. and Hwang, Harold Y. and Mundy, Julia A. and Muller, David A. and Goodge, Berit H.},
  title = {{Structural modifications in strain-engineered bilayer nickelate thin films}},
  journal = {Nature},
  volume = {653},
  pages = {76--82},
  year = {2026},
  doi = {10.1038/s41586-026-10446-2}
}

@article{XHChen2026,
  author = {Wang, Zhiwei and Wang, Zhengjie and Wang, Huiyu and Zhu, Mingyi and Shi, Mengzu and Huang, Zongyao and Li, Houpu and Ning, Shoucong and Tao, Jing and Wu, Tao and Chen, Xianhui},
  title = {{Electron-like high-temperature superconductivity induced by compressive strain in La$_2$PrNi$_2$O$_7$ thin films}},
  journal = {arXiv:2608.01295},
  year = {2026},
  eprint = {2608.01295},
  archivePrefix = {arXiv},
  primaryClass = {cond-mat.supr-con}
}

@article{RevModPhys.83.1057,
  author = {Qi, Xiao-Liang and Zhang, Shou-Cheng},
  title = {{Topological insulators and superconductors}},
  journal = {Rev. Mod. Phys.},
  volume = {83},
  pages = {1057--1110},
  year = {2011},
  doi = {10.1103/RevModPhys.83.1057}
}

@article{Alicea_2012,
  author = {Alicea, Jason},
  title = {{New directions in the pursuit of Majorana fermions in solid state systems}},
  journal = {Rep. Prog. Phys.},
  volume = {75},
  pages = {076501},
  year = {2012},
  doi = {10.1088/0034-4885/75/7/076501}
}

@article{10.1093/ptep/ptae065,
  author = {Tanaka, Yukio and Tamura, Shun and Cayao, Jorge},
  title = {{Theory of Majorana Zero Modes in Unconventional Superconductors}},
  journal = {Prog. Theor. Exp. Phys.},
  volume = {2024},
  pages = {08C105},
  year = {2024},
  doi = {10.1093/ptep/ptae065}
}

@article{PhysRevB.61.10267,
  author = {Read, N. and Green, Dmitry},
  title = {{Paired states of fermions in two dimensions with breaking of parity and time-reversal symmetries and the fractional quantum Hall effect}},
  journal = {Phys. Rev. B},
  volume = {61},
  pages = {10267--10297},
  year = {2000},
  doi = {10.1103/PhysRevB.61.10267}
}

@article{QIN20191207,
  author = {Qin, Shengshan and Hu, Lunhui and Wu, Xianxin and Dai, Xia and Fang, Chen and Zhang, Fu-Chun and Hu, Jiangping},
  title = {{Topological vortex phase transitions in iron-based superconductors}},
  journal = {Sci. Bull.},
  volume = {64},
  pages = {1207--1214},
  year = {2019},
  doi = {10.1016/j.scib.2019.07.011}
}

\end{document}